# From Weak to Strong Coupling: Quasi-BIC Metasurfaces for Mid-infrared Light-Matter Interactions

Shovasis Kumar Biswas[a]†, Wihan Adi[b]†, Aidana Beisenova[b], Samir Rosas[b], Eduardo Romero Arvelo[a], Filiz Yesilkoy[b]*

[a] Department of Electrical and Computer Engineering, University of Wisconsin-Madison Madison, WI 53706, USA

[b] Department of Biomedical Engineering, University of Wisconsin-Madison Madison, WI 53706, USA

*Email: filiz.yesilkoy@wisc.edu

† Authors contributed equally

**Abstract:**
Resonant metasurfaces present extraordinary subwavelength light trapping capabilities, which have been critical to the development of high-performance biochemical sensors and surface-enhanced spectroscopy techniques. To date, metasurface-enhanced light-matter interactions in the mid-infrared region have been primarily leveraged in the weak coupling regime. Nevertheless, the strong coupling regime, characterized by the hybrid light-matter states – polaritons, has not been fully explored. Specifically, metasurfaces made from low-loss dielectric resonators underpinned by the bound states in the continuum (BIC) are well suited for quantum coherent energy exchange between their high-Q open cavities and molecules' resonant transitions. This paper delves into the light-matter interactions in metasurface cavities generated by quasi-BIC resonances at mid-infrared frequencies. We thoroughly investigate the material and cavity parameters that define the boundaries between weak and strong coupling. Our findings underscore the transformative potential of dielectric metasurfaces to harness vibrational strong coupling for applications such as cavity polaritonic chemistry, modulation of chemical reactivity with light, and highly sensitive molecular spectroscopy.

1. Introduction

Photonic metasurfaces present tremendous opportunities in optics and photonics via efficient spatial and temporal control of light at ultrathin engineered interfaces. A significant development in this domain is the emergence of all-dielectric metasurfaces made from structured high-index dielectric or semiconductor materials with low energy dissipation[1]. Notably, dielectric metasurfaces can confine light intensely into the near-field hotspots, presenting remarkable opportunities for the subwavelength control of light-matter interactions[2–4]. Specifically, resonances that are based on the bound states in the continuum (BIC) light trapping mechanism can generate accessible and powerful photonic nanocavities on the metasurfaces[5,6]. Among various photonic structures that can support BICs, metasurfaces built using meta-atoms with broken in-plane inversion symmetry can substantially suppress the radiative losses, leading to sharp resonances associated with quasi-BIC modes (q-BICs)[7]. Across the electromagnetic spectrum, enhanced light-matter interactions supported by q-BIC modes have enabled bourgeoning applications in biochemical sensing[8–10], surface-enhanced infrared absorption spectroscopy (SEIRAS)[11], THz detection[12,13], nonlinear light generation[14,15], enhancing fluorescence[16], photoluminescence[17], and chiral optical response[18]. To date, metasurface applications have usually leveraged light-matter interactions in the weak coupling regime. Exciting technological prospects of quantum-coherent strong light-matter interactions using metasurface photonic cavities are yet to be fully explored.

In the strong coupling regime, the coupling strength between a cavity mode and a molecular transition exceeds the system's total energy dissipation rates. This leads to a recurrent coherent energy exchange between the cavity and molecular transition modes. The strongly coupled light-matter hybrid system is characterized by new energy states separated by the Rabi energy constant ($\hbar\Omega$). A frequency domain analysis can reveal the spectrum of hybrid light-matter states called polaritons, which are quasiparticles simultaneously associated with light and molecules' transitions.

A photonic cavity's resonance mode can be designed to strongly couple to electronic or vibrational transitions in a molecule, generating exciton-polaritons or vibro-polaritons, respectively[19]. Since its first demonstration by Yakovley et al. in 1975[20], strong exciton-photon coupling has been extensively studied using polymers[21], dye molecules[22], quantum dots[23], and transition metal dichalcogenides[24], etc., enabling the observation of some profound effects, including Bose–Einstein condensation[21], and superfluidity[25]. Vibrational strong coupling (VSC), on the other hand, was only recently experimentally demonstrated in 2015[26–28]. Since then, there has been a tremendous interest in VSC due to its exciting potential to modify the chemical reactivity of matter. However, achieving VSC is more challenging than strong coupling to electronic states, which are associated with high energy transitions (1-3 eV) and involve electrons delocalizing the entire molecule. Vibrational transitions, however, have lower energy (~ 30 – 500 meV) and are localized to specific bonds in a molecule. This highlights the critical need to develop superior light-harnessing cavities to enable investigations of VSC at the molecular level.

The key prerequisites for cavity-coupled VSC include low photon loss (high-Q or small linewidth) in both the cavity and material, strong resonance light confinement (small mode volumes), and ease of cavity resonance frequency tunability to specific vibrational bands. Additionally, open cavities that enable easy sample access to the photonic modes are crucial for broad photochemical investigations. The Fabry-Perot (FP) cavities with metallic mirrors are the most extensively used systems to explore VSC due to their easy fabrication, simple resonance tuning by gap distance control, robust resonances, and straightforward spectral dispersion analysis by angle-resolved measurements[26,28,29]. However, FP cavities are limited in supporting high-Q resonances due to metallic losses of mirrors, and their large mode volumes can only support collective strong coupling to a large number of molecules. Other viable routes to VSC include coupling molecular transitions to propagating surface plasmon polaritons (SPP)[30,31], and localized surface plasmons supported by nanoantenna with lower mode volumes than the FP cavities[32–36]. However, plasmonic resonances suffer from Ohmic losses inherent to metallic resonators, and the resultant high damping rates hinder their strong coupling to vibrational transitions. Recently, the strong light-localization capabilities of high-Q dielectric metasurface supporting q-BIC resonances have been employed for VSC[37–39]. However, significant gaps remain in our understanding of these high-Q metasurface nanocavities for VSC applications, underscoring the necessity of fundamental studies on the parameters needed to realize coherent energy exchange in q-BIC cavities in the mid-infrared (MIR) range.

In this work, we explore the dynamics of VSC between q-BIC resonances in the MIR over a large parameter space, including cavity Q-factor (linewidth), intrinsic molecular losses and transition dipole moment strengths, material thickness filling the metasurface cavity, and the coupling efficiency between cavity mode and the molecular transitions. In our numerical investigations, we focus on the light-matter interactions between germanium (Ge) q-BIC metasurface with the well-known zigzag design[7] and the absorption band associated with the C=O stretching of the polymethyl methacrylate (PMMA) molecule (Fig. 1). Our findings reveal the parameter-dependent transition from weak to strong coupling regimes in a coupled system. We specifically leverage the giant yet tunable, resonance lifetimes of the q-BIC modes to demonstrate their substantial effects on the coupled cavity-molecule system. We show that the maximum polariton lifetime can be achieved across various material systems by tuning the cavity photon lifetime and spectral position to a molecule's unique two-level vibrational resonance. Furthermore, our analysis delves into the practical aspects of spectral polariton detection within a coupled system, revealing that a smaller cavity linewidth facilitates easier separation of polariton peaks. Overall, our results reveal the promising capabilities of the powerful photonic nanocavities supported by the q-BIC dielectric metasurface for studying light-matter interactions while providing design guidance for VSC applications addressing the growing interest in the emerging polaritonic devices field.

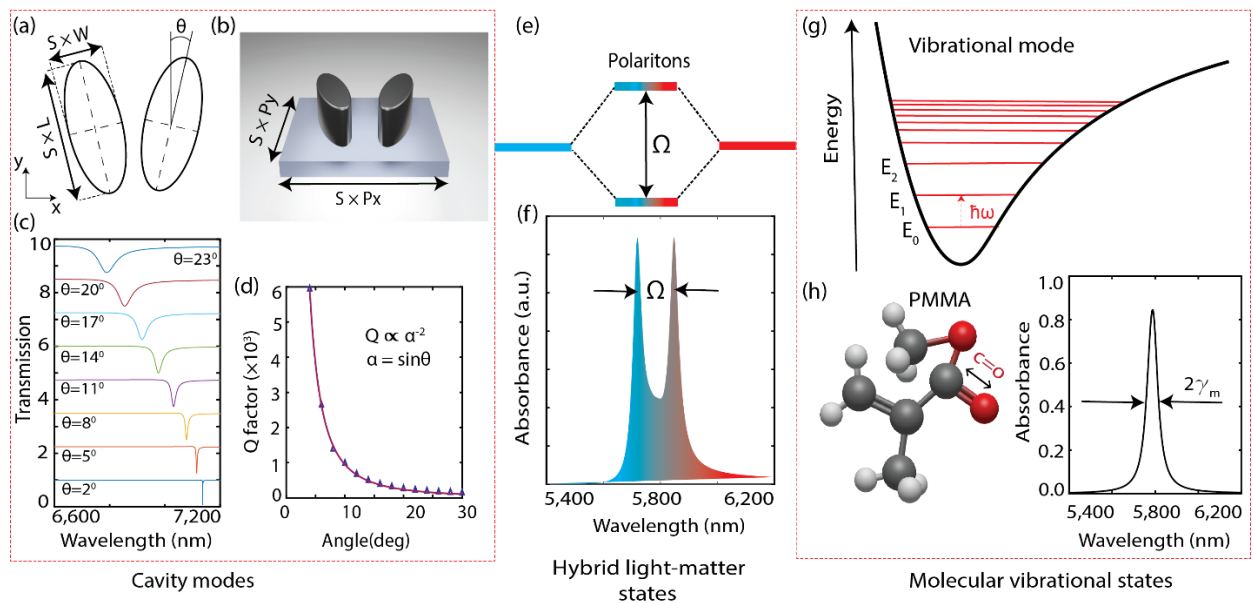

**Fig. 1: Overview of vibrational strong coupling between a photonic cavity mode and molecular vibrational states.** (a,b) Top-view schematic and a rendered image of the metaunit, consisting of two elliptical dielectric structures made of germanium (Ge). The metaunit lies in the x-y plane with period Px = 4050 nm, Py = 2340 nm, L = 1077 nm, and W = 500 nm, representing the long and short axes of the elliptical disks. Scaling parameter S and asymmetry parameter θ tune the resonance wavelength and Q-factor, respectively. (c) Numerically simulated metasurface transmission spectra as a function of the asymmetry parameter θ. (d) Q-factor of the q-BIC modes as a function of the asymmetry parameter θ, revealing adjustability of the photon lifetimes. The solid red line and blue triangles represent the fitted and calculated Q factors, respectively. (e) Vibro-polaritonic states distinctly separated by the Rabi splitting energy ($\Omega$). (f) Absorption spectra illustrating the formation of Rabi splitting under strong coupling. (g) A schematic energy level diagram depicting the molecular vibrational states. (h) The chemical structure of polymethyl methacrylate (PMMA) molecule and its absorption spectra corresponding to C=O stretching of its carboxyl group.

## 2. Results
### 2.1. Spectral characterization of VSC

Fig. 1 a and b depict schematics of the studied metasurface structure, consisting of a zigzag array of Ge elliptical rods on a $CaF_2$ substrate. In this metasurface design, an asymmetry is introduced by tilting the two rods in a metaunit oppositely by θ degrees from their long vertical axes, which align with y-axis at θ = 0 degress. Thus, the induced in-plane asymmetry disrupts the ideal BIC condition and opens coupling channels to excite the q-BIC modes with horizontally polarized and normally incident plane waves. Since Ge is a low-loss material in the MIR, the radiation losses, which are controlled by the tilting angle, θ, determine the cavity linewidth. Fig. 1c, d shows the typical inverse quadratic dependence of the cavity Q-factor of q-BIC modes to the asymmetry parameter, θ.

Moreover, spectral tunability is achieved via a dimension scaling factor, S. In this work, we performed electromagnetic simulations using a finite-element frequency-domain solver (CST Studio Suite 2023) with Floquet periodic boundary conditions to study the light-matter interactions on the q-BIC metasurface. The optical constants for the Ge and CaF$_2$ were taken from Ref. [40,41].

The coherent dynamics of a molecule-cavity system can be modeled as two harmonic oscillators interacting with a coupling strength of g. When the complex frequencies of molecule ($\omega_m - i\gamma_m$) and cavity ($\omega_c - i\gamma_c$) are considered to account for the damping processes, the eigenfrequencies of the coupled system can be stated as in equation 1.

$$\omega_\pm = \frac{\omega_c + \omega_m}{2} - i\frac{\gamma_c + \gamma_m}{2} \pm \sqrt{g^2 + \left(\delta + i\frac{(\gamma_c - \gamma_m)}{2}\right)^2} \quad (1)$$

where $\delta = \omega_c - \omega_m$ is the detuning between cavity resonance and molecular vibration modes, and $\gamma_c, \gamma_m$ are the cavity and material linewidths, respectively[42,43]. At non-zero detuning ($\delta \neq 0$), the Rabi splitting, defined as the difference between the eigenfrequencies, can be expressed as in equation 2.

$$\Omega = \omega_+ - \omega_- = 2\sqrt{g^2 + \left(\delta + i\frac{(\gamma_c - \gamma_m)}{2}\right)^2}. \quad (2)$$

$$\Omega_0 = \sqrt{4g^2 - (\gamma_c - \gamma_m)^2}. \quad (3)$$

At the zero detuning resonance case ($\delta = 0$), the difference between the eigenfrequencies becomes equation 3, which takes real values when $2g > |\gamma_c - \gamma_m|$. However, this theoretical condition does not guarantee observation of the two distinct polariton modes in practice. To observe strong coupling in an experimental system, the Rabi splitting frequency ($\Omega$) needs to exceed the polariton bandwidth, which is the average of the cavity and molecular decay rates (i.e. $\Omega > \frac{\gamma_c + \gamma_m}{2}$). Together, these two conditions impose the strong coupling criterion as $g > \sqrt{(\gamma_c^2 + \gamma_m^2)/2}$.

In this work, to study the resonant light-matter interactions, we spectrally sweep the q-BIC resonance by changing the S parameter through the molecular vibration mode of the PMMA molecule (Fig. 2). Fig. 2a demonstrates the impact of spectral detuning ($\delta$) within the coupled system. When $\delta$ is non-zero, the hybrid mode frequencies, $\omega_\pm$, exhibit asymmetry across the molecular vibration mode, and the anti-crossing feature appears (Fig. 2c), which is a requisite for the strong coupling. At zero detuning ($\delta = 0$), the coupled system is characterized by the Rabi splitting $\Omega_0 = \sqrt{4g^2 - (\gamma_c - \gamma_m)^2}$, which designates the quantitative coupling strength (Fig. 2b). In the following analysis, the Rabi splitting in each cavity-coupled material system is identified by, first, polynomial fitting of the simulated hybrid mode splitting values as a function of S, and then assigning the minimum of this polynomial as the value at zero detuning $\Omega_0$ (Fig. 2b).

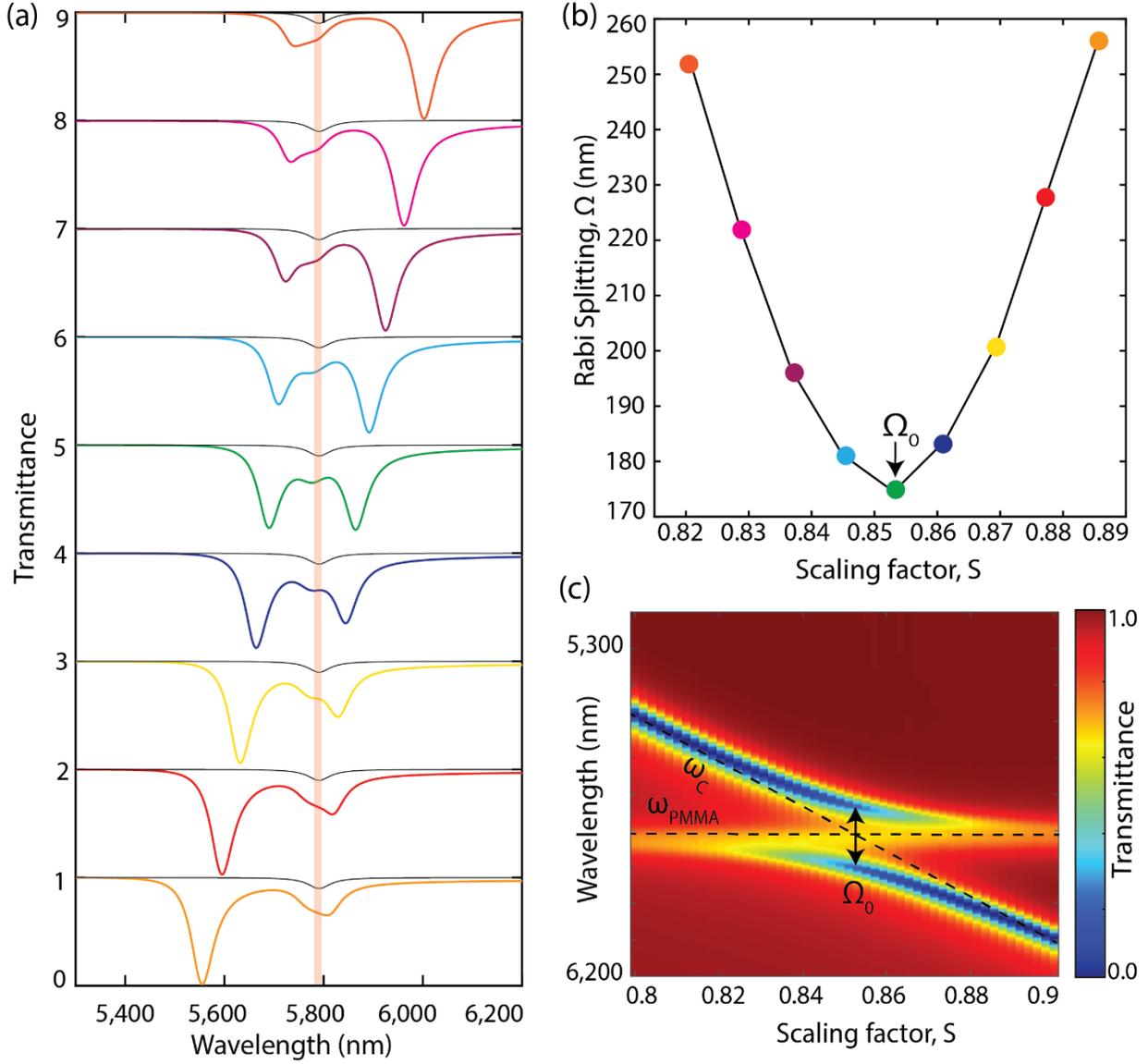

**Fig. 2: Spectral detuning between cavity resonance and molecular vibration transition mode**. (a) Numerically simulated transmission spectra of a coupled system at various detuning where q-BIC metasurface cavity (θ = 20 degrees) resonance is spectrally swept through 50 nm PMMA's C=O absorption peak at 5,780 nm resonance wavelength by varying the metasurface scaling factor, S. The transmission spectra of 50 nm PMMA layer without the metasurface is shown with the gray line. (b) Rabi splitting, $\Omega$ between the hybrid modes ($\omega_\pm$) as a function of metasurface scaling parameter, S, where $\Omega_0$ marks the splitting at the zero-detuning. (c) Transmission spectra of the coupled system, numerically calculated by varying the metasurface scaling factor, S, reveal the hybrid mode anti-crossing feature of the VSC. The uncoupled molecule's vibrational frequency, $\omega_{PMMA}$, and cavity resonance mode, $\omega_c$, are individually marked with black dashed lines.

## 2.2. Cavity linewidth and material thickness

The coupling constant between a single two-level vibrational transition system and a cavity is defined by $g_0 = \mu \cdot E_c$, where $\mu$ is molecule's transition dipole moment and $E_c$ is the cavity vacuum electric field. According to this relation, the cavities that can better localize and enhance the density of states in a cavity can more efficiently couple to molecular transition modes for a given molecule. Moreover, when more than N molecular modes couples to the same cavity resonance, the collective coherence between the molecular ensemble and the cavity increases to[44] $g = \sqrt{N}g_0$.

To evaluate the ensemble molecule (N) effect on the coupling strength, we calculated the Rabi splitting as a function of PMMA material thickness on a metasurface. By varying the PMMA layer thickness (t=3, 5, 10, 20, 30, and 50 nm), we controlled the total number of molecules (N) within a single cavity. Moreover, we repeated this study for three different q-BIC modes with varying cavity linewidths, $\gamma_c$ =8.2, 30, and 56.7 nm, achieved by increasing the tilting angle, $\theta$ =10, 20, and 30 degrees, respectively (Fig. 3). On the right column of Fig. 3a-c, we plot the Rabi splitting values calculated for each material thickness. The findings depicted in Fig. 3 demonstrate the $\sqrt{N}$ relation between the Rabi splitting and the PMMA layer thickness, as expected.

Notably, the coupling strength signified by the Rabi splitting versus the material thickness relation is not identical in cavities with different photon damping characteristics (Fig. 3 a-c). We calculated the strong coupling criteria for each cavity linewidth, $\gamma_c$ =8.2, 30, and 56.7 nm, as $\Omega_0 > 38.2$ nm, 60.0 nm, and 86.7 nm, respectively. Our investigation in Fig. 3 revealed that the minimum material thickness required to satisfy the strong coupling condition increases with the cavity linewidth, associated with the total cavity photon losses. The simulation results in Fig. 3a shows that the strong coupling condition can be satisfied with as thin as 3 nm PMMA when the q-BIC mode is sharp with 8.2 nm linewidth. However, an increase in material thickness is necessary to achieve strong coupling when the cavity linewidth increases from 8.2 nm to 30.0 nm and 60.0 nm. With the higher loss cavities (larger $\gamma_c$), when the total molecule number (N) is low, the ensemble coupling strength (g) is not sufficient for strong coupling, and the light-matter interactions are characterized under the weak coupling condition (g $\leq \sqrt{(\gamma_c^2 + \gamma_m^2)/2}$ ). The weak coupling condition is usually characterized by a resonance peak damping and quantified by the photonic resonance amplitude decrease in the presence of absorbing molecules[11].

However, the transition between the weak and strong coupling is usually not abrupt. Therefore, here, we define a new "transition coupling region", where the polaritons can be observed in the spectrum as the Rabi frequency has a real but small value ($|\Omega_0| > 0$), even though the strong coupling criterion is not met. For example, in Fig. 3a, the presence of hybridized mode splitting in the transmission spectra (lower subfigure) is observable for 3 nm material thickness. This cavity material system doesn't satisfy the strong coupling condition; however, the spectrum diverges from a simple mode damping scheme, which is a characteristic of weak coupling. Thus, we define the new transition region condition as $|\Omega_0| > 0 \; and \; g \leq \sqrt{(\gamma_c^2 + \gamma_m^2)/2}$ ).

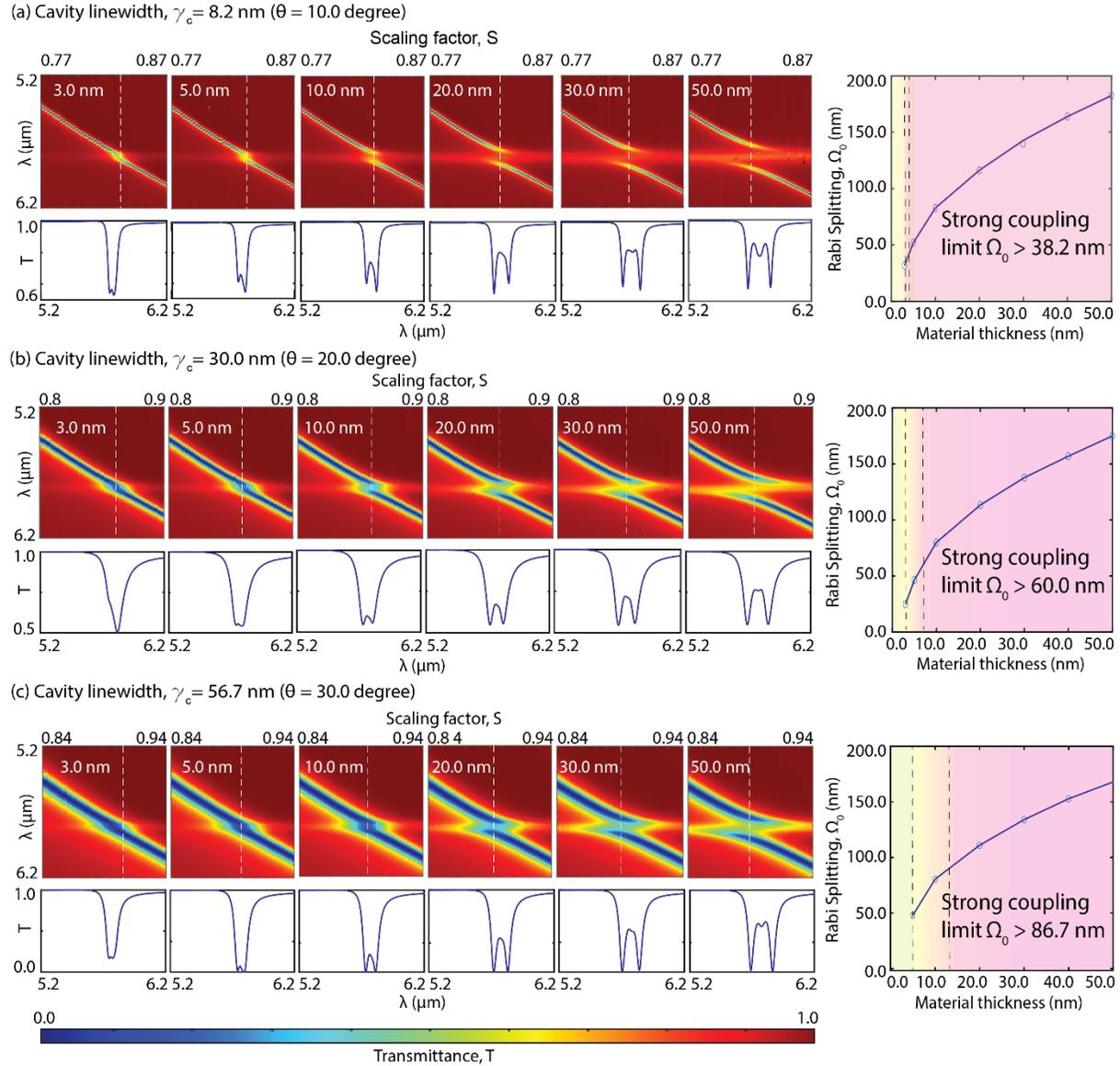

**Fig. 3: Effects of cavity linewidth and material thickness on the coupled system.** The material thickness effect is shown for three different cavity linewidths with (a) $\gamma_c$=8.2 nm achieved by θ=10 degrees, (b) $\gamma_c$=30 nm achieved by θ=20 degrees and (c) $\gamma_c$=56.7 nm achieved by θ=30 degrees. For each cavity linewidth, the transmission spectral maps as a function of scaling factor (i.e. various detuning), S, are presented for different PMMA layer thicknesses (3.0 nm, 5.0 nm, 10.0 nm, 20.0 nm, 30.0 nm, and 50.0 nm), showing the anti-crossing patterns. Under each spectral map, a single transmission spectrum, T, at zero detuning point is shown. The white dashed lines indicate the zero detuning points for each coupled system. On the right column, plots show the Rabi splitting at the zero detuning, $\Omega_0$, as a function of material thickness. From (a) to (c), as the metasurface cavity linewidth increases, the weak-to-strong coupling transition shifts to higher material thicknesses, indicating the requirement of a larger number of collectively coupled molecules (N) for VSC to compensate for the cavity's

photon loss. The newly defined transition region is demonstrated with yellow and yellow-to-pink transition colors, and the strong coupling regime is indicated by pink color.

## 2.3. Material parameters

To reveal the dependency of the VSC on the molecular vibrational resonance properties, we studied various artificial materials, which are derived from the C=O bond stretching resonance in the PMMA molecule. The complex permittivity of these materials is calculated using the Lorentz model for dispersion in dielectrics[45]:

$$\varepsilon(f) = \varepsilon_\infty + \frac{\Delta\varepsilon f_0^2}{f_0^2 - 2jf\gamma_m - f^2}$$

Where $\varepsilon_\infty$ is the background relative permittivity, $f_0$ is the Lorentz resonance frequency, $\Delta\varepsilon$ is the absorbance strength coefficient and $\gamma_m$ the Lorentz damping rate of the molecule. Some realistic parameter values used to model the artificial materials are provided in Tables 1 and 2. Coupling the designated materials to the q-BIC cavities allowed us to study correlations between Rabi energy exchange rate and materials' absorption strength and nonradiative losses. In IR absorption spectroscopy, the amplitude of absorption depends on the polarity of the bond, as well as the concentration of the bonds in a molecule. Therefore, the same functional group can have different absorption strengths ($\Delta\varepsilon$) depending on the molecule. Moreover, when collecting IR spectra from a compound in liquid or solid phase, many individual molecules interact with each other. Thus, individual bond absorptions occur at varying frequencies, broadening the absorption peaks at different levels. The dephasing due to electron-photon interactions can be reduced by cooling the sample to cryogenic temperatures, which can reduce ($\gamma_m$). In our strong coupling investigations, we consider these realistic material properties to interrogate each designated material's cavity-coupled interactions.

Table 1: Lorentz oscillator parameters used to design three artificial materials with different absorption strength $\Delta\varepsilon$ with the same damping rates $\gamma_m$. The imaginary parts of the permittivity for the three designated materials are plotted around the resonance frequency ($f_0$), showing the characteristic Lorentzian spectral line shape with fixed linewidth.

| # | $\varepsilon_\infty$ | $\gamma_m$ (THz) | $f_0$ (THz) | $\Delta\varepsilon$ |
|---|---|---|---|---|
| 1 | 2.198 | 0.235 | 51.77 | 0.025 |
| 2 | 2.198 | 0.235 | 51.77 | 0.050 |
| 3 | 2.198 | 0.235 | 51.77 | 0.070 |

$\gamma_{m1}$= 30.0 nm
$\gamma_{m2}$= 30.0 nm
$\gamma_{m3}$= 30.0 nm
Abs. (a.u.)
λ (μm)

Table 2: Lorentz oscillator parameters used to design three artificial materials with different damping rates $\gamma_m$ while keeping their absorption strengths $\Delta\varepsilon$ the same. The imaginary parts of the permittivity for the three designated materials are plotted around the resonance frequency ($f_0$), showing the characteristic Lorentzian spectral line shape with varying linewidth.

| # | $\varepsilon_\infty$ | $\gamma_m$ (THz) | $f_0$ (THz) | $\Delta\varepsilon$ | |
|---|---|---|---|---|---|
| 4 | 2.198 | 0.10 | 51.77 | 0.009 | 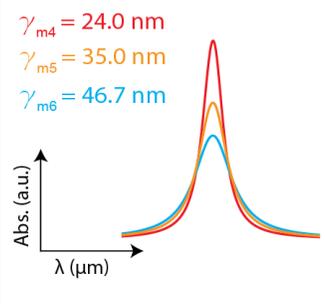 |
| 5 | 2.198 | 0.15 | 51.77 | 0.009 | |
| 6 | 2.198 | 0.20 | 51.77 | 0.009 | |

In Fig. 4, we present our findings on the coupling strength as a function of material properties and the cavity linewidth. Fig. 4a shows the Rabi splitting for three materials with different absorption strengths ($\Delta\varepsilon$) and the same damping rates ($\gamma_m$) as a function of the cavity linewidths ($\gamma_c$=10 – 50 nm) that they couple to. Since the absorbance peak is associated with the molecule's transition dipole moment ($\mu$), which determines the coupling constant ($g = \sqrt{N}\,(\mu \cdot E_c)$), the total Rabi splitting at zero detuning $\Omega_0$ (equation 3) increases with the material absorbance strength. Moreover, when the damping rates of the material and the cavity are equal ($\gamma_c = \gamma_m$), the critical coupling condition is met, maximizing the strong coupling efficiency at $\Omega_0 = 2g$. In Fig. 4b, the Rabi splitting is depicted for three materials characterized by varying damping rates ($\gamma_m$) but uniform absorption strengths ($\Delta\varepsilon$), as a function of the cavity linewidths ($\gamma_c$=5 – 30 nm) to which they are coupled. The material with the smallest damping rate ($\gamma_{m4}$) exhibits the highest absorption peak among the three materials, because of the lower loss.

Further, we analyzed the cavity linewidth effects on the separability of polariton peaks in a strongly coupled system considering the materials defined in Tables 1 and 2. Here we defined polariton separability as $\frac{\Omega_0}{(\gamma_c + \gamma_m)/2}$, which measures how easily polaritons can be identified in an experimental spectral analysis. As expected, narrower material linewidths facilitate better separability in any material system (Fig. 4 c,d). The polariton spectra presented in the insets illustrates our findings. Moreover, our investigation revealed that larger absorption strength facilitates the polariton peak separation due to higher $\Omega_0$ (Fig. 4c). Furthermore, materials with the smallest non-radiative losses (small $\gamma_m$), maximizes the polariton separability (Fig. 4d). This intriguing observation emphasizes the pivotal role played by material characteristics, particularly the linewidth and absorption strength, on the separability of polariton peaks within a coupled system. The interplay between the material and the cavity properties sheds light on the multifaceted attributes of identifying polariton peaks in an experimental investigation.

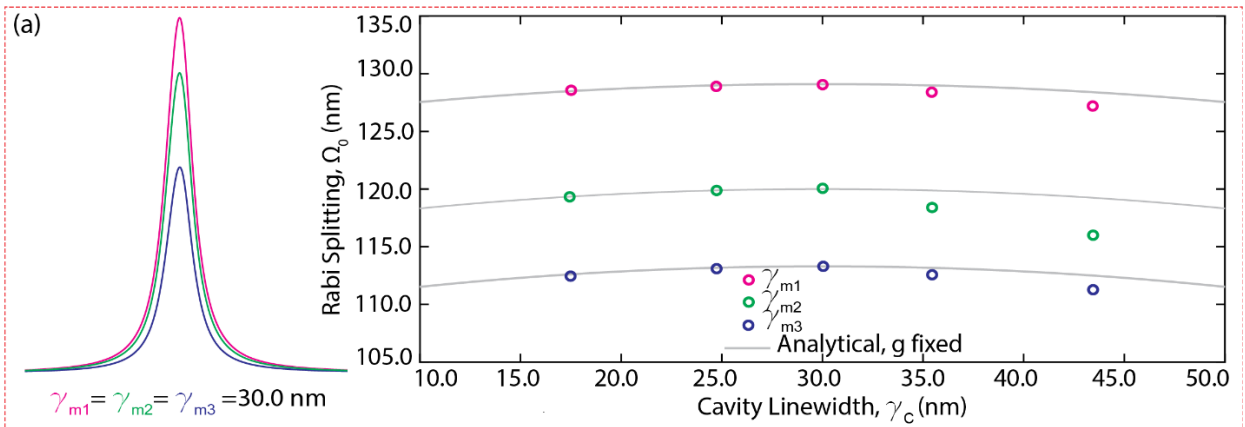

(a)

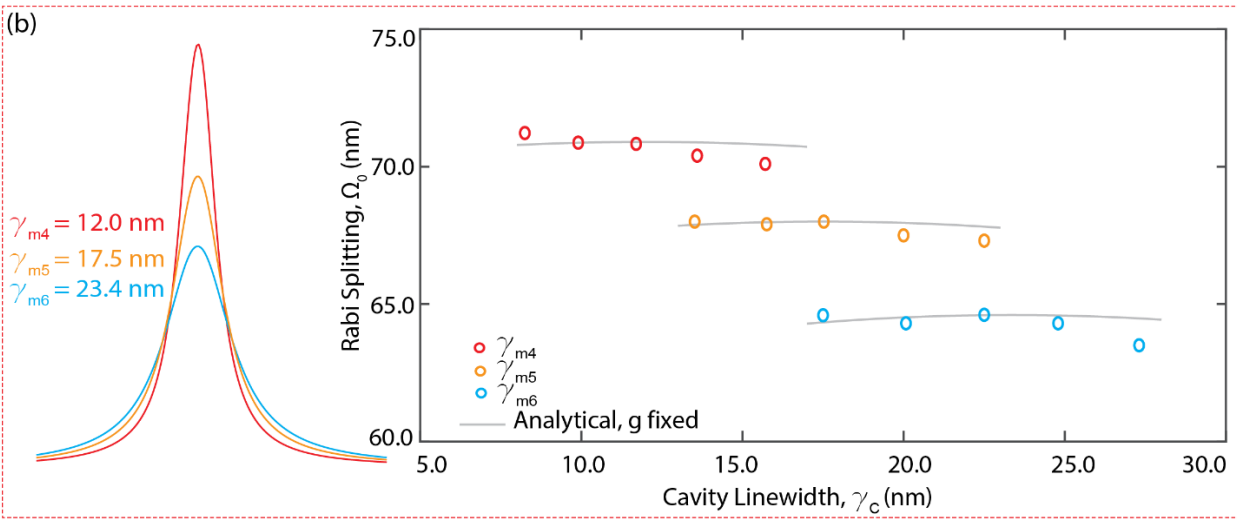

(b)

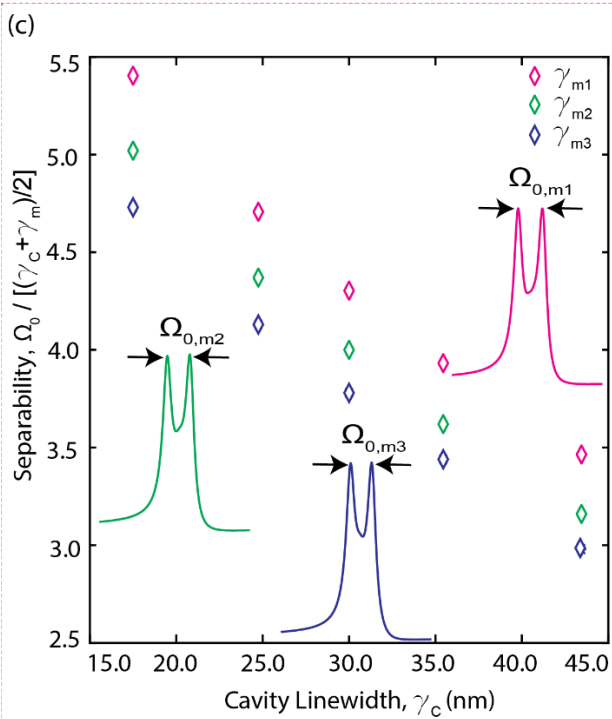

(c)

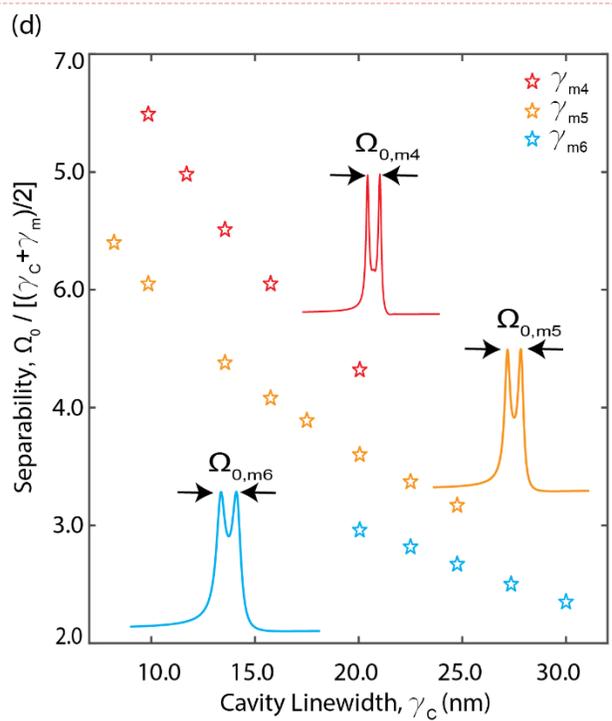

(d)

**Fig. 4: Effects of material loss and absorption strength on the coupled system.** (a) Numerically calculated $\Omega_0$ as a function of cavity linewidth, $\gamma_c$, for three different materials with the same absorption linewidth but different absorption strength. (b) Numerically calculated $\Omega_0$ as a function of cavity linewidth, $\gamma_c$ for three different materials with different damping rates but same absorption strength. The solid gray lines in (a) and (b) denote the analytical $\Omega_0$ calculated with a fixed coupling strength, g (equation 3). Analysis of polariton peak separability as a function of cavity linewidth, $\gamma_c$, observed across materials used in (a) and (b) are shown in (c) and (d), respectively. The polariton spectra with Rabi splitting at zero detuning $\Omega_{0,m1}$ to $\Omega_{0,m6}$ correspond to the materials $\gamma_{m1}$ to $\gamma_{m6}$ are also shown.

## 2.4. Cavity-material coupling efficiency.

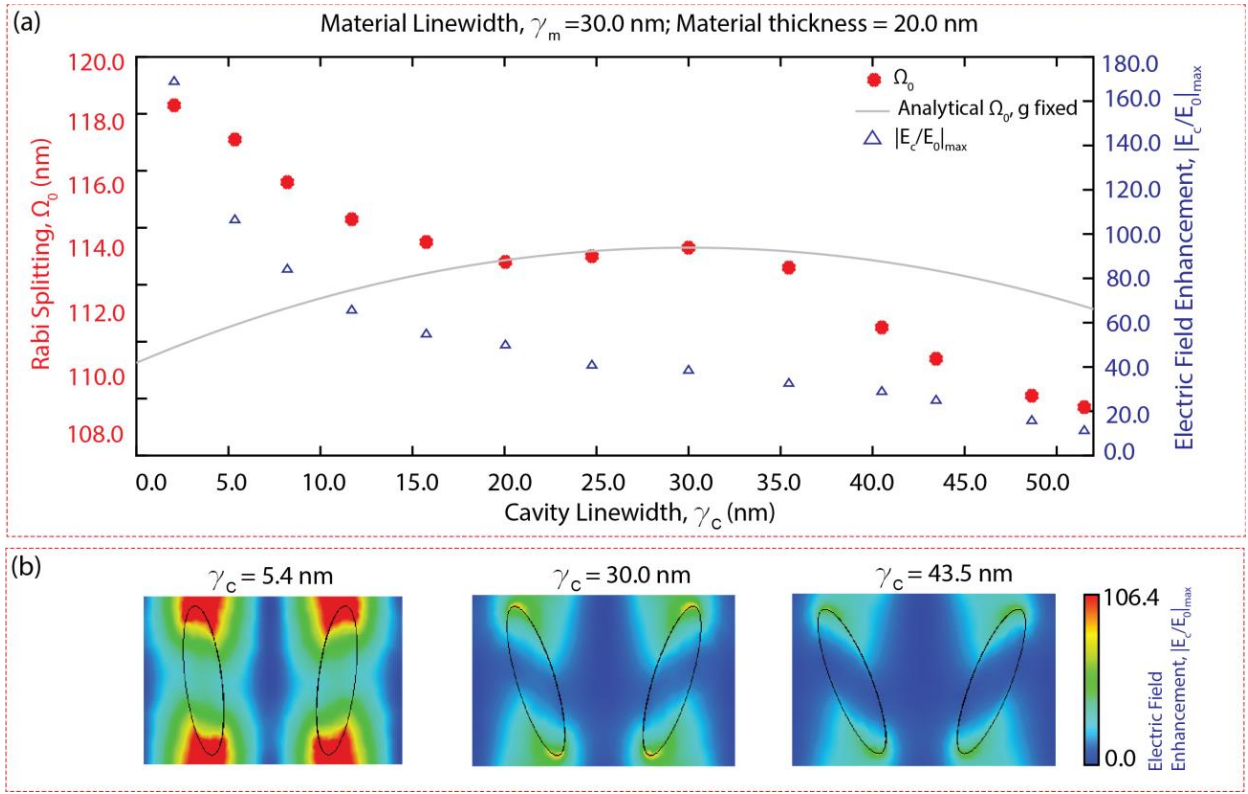

**Fig. 5: Critical strong coupling and cavity loss effects on the light-matter interactions.** (a) For the same material type and thickness, the calculated $\Omega_0$ (denoted by red asterisk) is plotted as a function of cavity linewidth, $\gamma_c$. The grey line shows the analytically calculated Rabi splitting as a function of cavity linewidth, assuming a constant coupling strength, g (equation 3). The parabolic dependency of $\Omega_0$ on the cavity linewidth displays the effects of critical coupling and loss mismatch between the cavity and the material. The discrepancy between the simulation (red asterisk) vs. analytical (grey line) underscores that the cavity linewidth affects the coupling strength, g. The maximum field enhancement indicated by the blue triangles is plotted as a function of cavity linewidth, $\gamma_c$. As the cavity loss decreases, absolute E-field enhancement and the coupling strength, g, increase, boosting the $\Omega_0$. (b) The electric field enhancement maps for three different metasurface cavities are depicted with reference to the same color bar, corroborating that the high Q-factor cavities are more suitable for VSC.

Furthermore, we investigated the coupling strength within the hybrid system by adjusting the Q-factor of the q-BIC cavity mode. In Fig. 5(a), we present $\Omega_0$ (red asterisks) values calculated by simulating a 20 nm thick PMMA with a linewidth of $\gamma_m$=30 nm on a q-BIC metasurface whose cavity linewidth was swept from $\gamma_m$=2 to 52 nm by varying the tilting angle, $\theta$. The grey line shows the analytically calculated equation 3, as a function of the cavity linewidth ($\gamma_c$), assuming a constant material damping ($\gamma_m$) and constant coupling strength, g. The parabolic dependency of $\Omega_0$ on the cavity loss displays how the coupling efficiency reaches a local maximum when the critical coupling ($\gamma_c = \gamma_m$) condition is satisfied and how the loss mismatch between the cavity and the material decreases the coupling efficiency. The discrepancy between the simulation (red asterisk) vs. analytical (grey line) results underscores that cavity linewidth, $\gamma_c$, affects the coupling strength, $g = \sqrt{N}\,(\mu \cdot E_c)$, by changing the field enhancement (blue triangle). In Fig. 5b, we present the absolute electric field enhancement maps ($|E_c/E_0|$) within a metaunit for three q-BIC metasurfaces whose tilting angle is adjusted to yield $\gamma_c$ of 5.4, 30 and 43.5 nm. As the maps show, the cavity field becomes stronger as the cavity linewidth decreases (i.e., metasurface Q-factor increases). Particularly, in Fig. 5a, we present maximum field enhancement (blue triangles) as a function of cavity linewidth, $\gamma_c$. As the cavity loss decreases, E-field enhancement increases, and thus, the coupling strength, g, boosts $\Omega_0$. This interplay between the cavity quality factor and photon loss rate mismatch between the cavity and material provides crucial insights into the dynamics governing the coupling behaviors in a hybrid system.

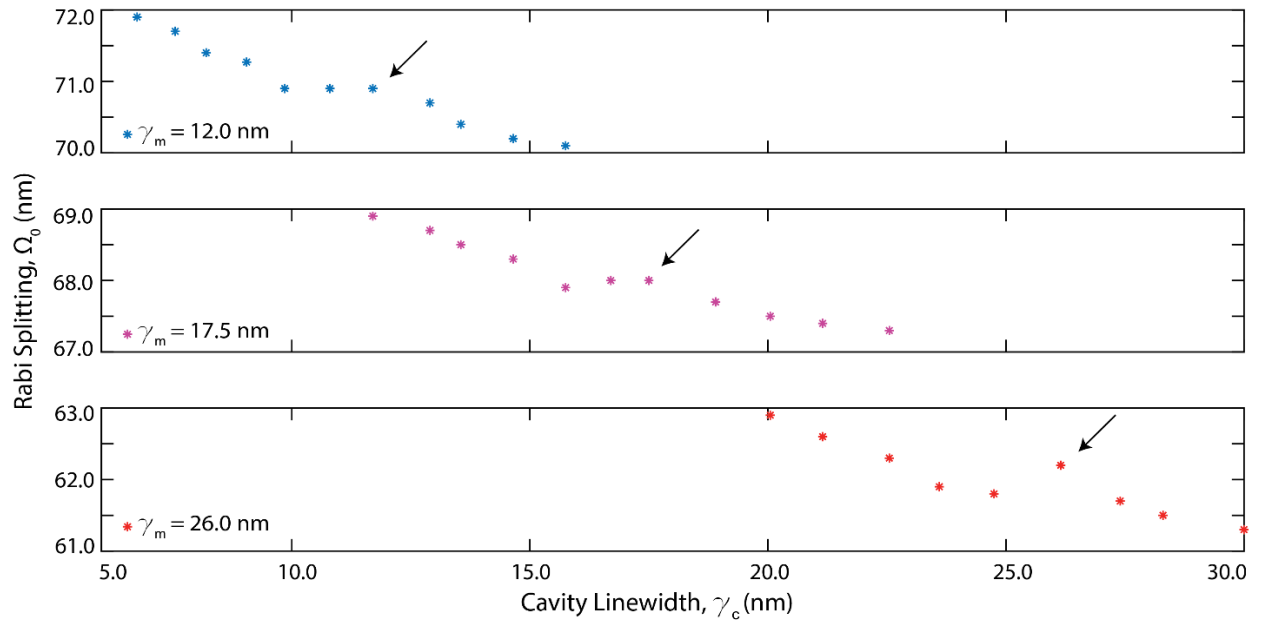

**Fig. 6: Critical strong coupling and material loss effects on the light-matter interactions.** Rabi splitting as a function of cavity linewidth plotted for three different materials. As the material loss decreases ($\gamma_m = 26\ nm$ to $\gamma_m = 12\ nm$), critical strong coupling shift toward the high-Q cavity region, causing an increase in coupling strength, g, and boosting the $\Omega_0$. Further boost of $\Omega_0$ is also achieved at critical coupling as shown with the black arrows.

Finally, we explored the critical coupling effects for different material-cavity systems. Fig. 6, shows the Rabi splitting ($\Omega_0$) variation as a function of cavity linewidth ($\gamma_c$=5 to 30 nm) for three different materials ($\gamma_m$=12.0 nm, 17.5 nm, and 26.0 nm). Our results reveal that it is favorable to work with high-Q cavities to achieve strong coupling independent of the material's vibrational transition characteristics. However, when real-world metasurface fabrication and intrinsic material losses limit the maximum achievable resonance Q-factors, satisfying the critical coupling condition by cavity-material loss matching can also enhance the coupling strength (as indicated by the black arrows). This effect is specifically pronounced in materials with large absorption linewidths (see Fig. 6 bottom). Overall, Fig. 6 unveils how the Q-factor tunability of q-BIC cavities supported by resonant dielectric metasurface is important for VSC applications.

## 3. Discussion

This work was initially motivated by the two published prominent reports using similar dielectric metasurface supporting the q-BIC resonance modes based on the zigzag geometry. In the work by Tittl et al., the light-matter interactions in the weak coupling regime were leveraged for biochemical sensing of proteins and polymer layers[11]. Meanwhile, Sun et al. experimentally demonstrated strong vibrational coupling with PMMA layers using similar metasurfaces[37]. Therefore, to elucidate the underlying physical parameters that determine the light-matter interaction regime, in this work, we extensively investigated the resonant cavity parameters of the zigzag metasurface design, such as the Q-factor (tuned by the symmetry-breaking factor, θ), and spectral position of the resonance (tuned by the scaling factor, S). We also considered material parameters, such as the absorption bandwidth related to the molecules' dephasing and internal losses, transition dipole moment strength, and material thickness filling the cavity void. Our findings unveiled that as the metasurface Q-factor increases, the minimum number of molecules in a single cavity (material layer thickness on a metasurface) requirement for VSC decreases. This collective molecule-cavity coupling effect explains the weak coupling regime in Tittl et al., where thin protein monolayers were measured, and the strong coupling regime in Sun et al., where 10s of nm thick polymer layers were measured[37].

Moreover, previous VSC investigations mainly used photonic cavities where the cavity photon loss rates were dominantly imposed by the materials, specifically metals' Ohmic losses, such as in the FP and plasmonic cavities. Therefore, the effects of the photonic cavity's Q-factor on the light-matter interactions were mostly unexplored in the MIR. Recently, all-dielectric metasurface introduced unprecedented efficiency in photon trapping into nanocavities, diminishing the limitations of material loss. Notably, metasurface supporting the q-BIC modes enabled the precise control of cavity photon loss rates into radiation channels via the geometrically tunable in-plane asymmetry parameters. Thus, in this work, we disclosed the benefits of cavity Q-factor tunability on the light-matter coupling efficiency.

As a rule of thumb, when using dielectric metasurface for VSC applications, the first goal must be to minimize the metasurface material and radiation losses to maximize the cavity Q-factor, which increases the resonance field enhancement and thus the coupling constant, g. However, in real-world applications, metasurface fabrication imperfections, together with non-zero material loss rates, constrain the practically achievable Q-factors. Then, a second important criterion for VSC and efficient light-matter interactions is the material and cavity loss matching. Uniquely, dielectric metasurface can enable critical strong coupling conditions via their Q-factor tunability.

Generally, strong coupling is defined based on the Rabi splitting exceeding the system loss rate, enabling coherent energy exchange between the cavity and the molecule. However, the literature has been ambiguous on the quantitative limits of the strong coupling regime. In this work, we adhered to the condition where the exchange rate must exceed the average loss of the two coupled oscillators ($\Omega_0 > (\frac{\gamma_c + \gamma_m}{2})$). Others have proposed a strong coupling criterion where Rabi energy exchange exceeds the larger of the two loss rates ($\Omega_0 > \gamma_c, \gamma_0$). While the polaritons form when $|\Omega_0| > 0$, where $\Omega_0$ is small and polariton linewidth is large, spectroscopic interrogation of hybrid states depends on the experimental optical measurement limitations. Thus, in this work, we named this ambiguous condition as a "transition region" where hybrid modes are present ($|\Omega_0| > 0$), but strong coupling condition is not strictly satisfied ($\Omega_0 \leq (\frac{\gamma_c + \gamma_m}{2})$). For example, our findings revealed that materials with narrower linewidths and higher absorption peaks coupled to high-Q cavities enable clearer spectral peak separability. Usually, by leveraging the collective coupling (high N), Rabi splitting can be increased to bring the system to the strong coupling regime for spectral investigations. Exceptionally, our findings lay the compelling research question of whether transition region can be the zone where the single or low counts of molecules' strong coupling to a cavity can be measured.

## 4. Conclusions

In this work, we used CST simulation tools and coupled oscillator theory to numerically and analytically explore the cavity-coupled light-matter interactions on q-BIC metasurface. We comprehensively outlined the parameter space vital for the understanding of the coupling between resonant cavity modes and resonant vibrational transitions in molecules. We leveraged the geometrical Q-factor and spectral position tunability of the q-BIC resonances to identify the weak and strong coupling regimes, providing a deeper insight into the interaction dynamics within the coupled system. Our findings indicate that, to satisfy the VSC condition in lossy cavities, collective mode coupling to a metasurface resonance is necessary, and minimum number of required molecules for VSC increases as the cavity Q-factor decreases. Notably, we introduced a new transition light-matter coupling zone, which falls in between the weak and strong coupling extents, where polaritons form, but their linewidth prohibits their spectral identification. Moreover, we studied molecules' transition dipole strength, absorption bandwidth correlated to their internal loss, and the total number of molecules coupled to a single cavity, to gauge their individual effects on the light-matter interactions. Our results can provide clear guidelines for designing optimal dielectric metasurface for VSC applications using a broad range of molecules' vibrational transitions. Based on our comprehensive study, we anticipate that the photonic cavities generated by dielectric metasurface can make a huge impact in the emerging field of polaritonic chemistry, where ground states and chemical reactivity of molecules can be modified.

**Conflict of Interest**

All other authors declare that they have no competing interests.

**Acknowledgement**

F.Y. acknowledges financial support from the National Institutes of Health (grant no. R21EB034411).

**References**

(1) Koshelev, K.; Kivshar, Y. Dielectric Resonant Metaphotonics. *ACS Photonics* **2021**, *8* (1), 102–112. https://doi.org/10.1021/acsphotonics.0c01315.
(2) Krasnok, A.; Caldarola, M.; Bonod, N.; Alú, A. Spectroscopy and Biosensing with Optically Resonant Dielectric Nanostructures. *Advanced Optical Materials* **2018**, *6* (5), 1701094. https://doi.org/10.1002/adom.201701094.
(3) Kruk, S.; Kivshar, Y. Functional Meta-Optics and Nanophotonics Governed by Mie Resonances. *ACS Photonics* **2017**, *4* (11), 2638–2649. https://doi.org/10.1021/acsphotonics.7b01038.
(4) Yang, S.; He, M.; Hong, C.; Caldwell, J. D.; Ndukaife, J. C. Engineering Electromagnetic Field Distribution and Resonance Quality Factor Using Slotted Quasi-BIC Metasurfaces. *Nano Lett.* **2022**, *22* (20), 8060–8067. https://doi.org/10.1021/acs.nanolett.2c01919.
(5) Hsu, C. W.; Zhen, B.; Stone, A. D.; Joannopoulos, J. D.; Soljačić, M. Bound States in the Continuum. *Nat Rev Mater* **2016**, *1* (9), 16048. https://doi.org/10.1038/natrevmats.2016.48.
(6) Koshelev, K. L.; Sadrieva, Z. F.; Shcherbakov, A. A.; Kivshar, Y. S.; Bogdanov, A. A. Bound States in the Continuum in Photonic Structures. *Physics-Uspekhi* **2023**, *66* (5), 494–517.
(7) Koshelev, K.; Lepeshov, S.; Liu, M.; Bogdanov, A.; Kivshar, Y. Asymmetric Metasurfaces with High- Q Resonances Governed by Bound States in the Continuum. *Phys. Rev. Lett.* **2018**, *121* (19), 193903. https://doi.org/10.1103/PhysRevLett.121.193903.
(8) Shi, Y.; Wu, Y.; Chin, L. K.; Li, Z.; Liu, J.; Chen, M. K.; Wang, S.; Zhang, Y.; Liu, P. Y.; Zhou, X.; Cai, H.; Jin, W.; Yu, Y.; Yu, R.; Huang, W.; Yap, P. H.; Xiao, L.; Ser, W.; Nguyen, T. T. B.; Lin, Y.-T.; Wu, P. C.; Liao, J.; Wang, F.; Chan, C. T.; Kivshar, Y.; Tsai, D. P.; Liu, A. Q. Multifunctional Virus Manipulation with Large-Scale Arrays of All-Dielectric Resonant Nanocavities. *Laser & Photonics Reviews* **2022**, *16* (5), 2100197. https://doi.org/10.1002/lpor.202100197.
(9) Yesilkoy, F.; Arvelo, E. R.; Jahani, Y.; Liu, M.; Tittl, A.; Cevher, V.; Kivshar, Y.; Altug, H. Ultrasensitive Hyperspectral Imaging and Biodetection Enabled by Dielectric Metasurfaces. *Nat. Photonics* **2019**, *13* (6), 390–396. https://doi.org/10.1038/s41566-019-0394-6.
(10) Romano, S.; Mangini, M.; Penzo, E.; Cabrini, S.; De Luca, A. C.; Rendina, I.; Mocella, V.; Zito, G. Ultrasensitive Surface Refractive Index Imaging Based on Quasi-Bound States in the Continuum. *ACS Nano* **2020**, *14* (11), 15417–15427. https://doi.org/10.1021/acsnano.0c06050.

(11) Tittl, A.; Leitis, A.; Liu, M.; Yesilkoy, F.; Choi, D.-Y.; Neshev, D. N.; Kivshar, Y. S.; Altug, H. Imaging-Based Molecular Barcoding with Pixelated Dielectric Metasurfaces. *Science* **2018**, *360* (6393), 1105–1109. https://doi.org/10.1126/science.aas9768.

(12) Han, S.; Pitchappa, P.; Wang, W.; Srivastava, Y. K.; Rybin, M. V.; Singh, R. Extended Bound States in the Continuum with Symmetry-Broken Terahertz Dielectric Metasurfaces. *Advanced Optical Materials* **2021**, *9* (7), 2002001. https://doi.org/10.1002/adom.202002001.

(13) Lou, J.; Jiao, Y.; Yang, R.; Huang, Y.; Xu, X.; Zhang, L.; Ma, Z.; Yu, Y.; Peng, W.; Yuan, Y.; Zhong, Y.; Li, S.; Yan, Y.; Zhang, F.; Liang, J.; Du, X.; Chang, C.; Qiu, C.-W. Calibration-Free, High-Precision, and Robust Terahertz Ultrafast Metasurfaces for Monitoring Gastric Cancers. *Proceedings of the National Academy of Sciences* **2022**, *119* (43), e2209218119. https://doi.org/10.1073/pnas.2209218119.

(14) Zograf, G.; Koshelev, K.; Zalogina, A.; Korolev, V.; Hollinger, R.; Choi, D.-Y.; Zuerch, M.; Spielmann, C.; Luther-Davies, B.; Kartashov, D.; Makarov, S. V.; Kruk, S. S.; Kivshar, Y. High-Harmonic Generation from Resonant Dielectric Metasurfaces Empowered by Bound States in the Continuum. *ACS Photonics* **2022**, *9* (2), 567–574. https://doi.org/10.1021/acsphotonics.1c01511.

(15) Liu, Z.; Wang, J.; Chen, B.; Wei, Y.; Liu, W.; Liu, J. Giant Enhancement of Continuous Wave Second Harmonic Generation from Few-Layer GaSe Coupled to High-Q Quasi Bound States in the Continuum. *Nano Lett.* **2021**, *21* (17), 7405–7410. https://doi.org/10.1021/acs.nanolett.1c01975.

(16) Liu, L.; Wang, R.; Sun, Y.; Jin, Y.; Wu, A. Fluorescence Enhancement of PbS Colloidal Quantum Dots from Silicon Metasurfaces Sustaining Bound States in the Continuum. *Nanophotonics* **2023**, *12* (15), 3159–3164. https://doi.org/10.1515/nanoph-2023-0195.

(17) Lin, Y.-T.; Hassanfiroozi, A.; Jiang, W.-R.; Liao, M.-Y.; Lee, W.-J.; Wu, P. C. Photoluminescence Enhancement with All-Dielectric Coherent Metasurfaces. *Nanophotonics* **2022**, *11* (11), 2701–2709. https://doi.org/10.1515/nanoph-2021-0640.

(18) Shi, T.; Deng, Z.-L.; Geng, G.; Zeng, X.; Zeng, Y.; Hu, G.; Overvig, A.; Li, J.; Qiu, C.-W.; Alù, A.; Kivshar, Y. S.; Li, X. Planar Chiral Metasurfaces with Maximal and Tunable Chiroptical Response Driven by Bound States in the Continuum. *Nat Commun* **2022**, *13* (1), 4111. https://doi.org/10.1038/s41467-022-31877-1.

(19) Kim, Y.; Barulin, A.; Kim, S.; Lee, L. P.; Kim, I. Recent Advances in Quantum Nanophotonics: Plexcitonic and Vibro-Polaritonic Strong Coupling and Its Biomedical and Chemical Applications. *Nanophotonics* **2023**, *12* (3), 413–439. https://doi.org/10.1515/nanoph-2022-0542.

(20) Yakovlev, V. A.; Nazin, V. G.; Zhizhin, G. N. The Surface Polariton Splitting Due to Thin Surface Film LO Vibrations. *Optics Communications* **1975**, *15* (2), 293–295. https://doi.org/10.1016/0030-4018(75)90306-5.

(21) Plumhof, J. D.; Stöferle, T.; Mai, L.; Scherf, U.; Mahrt, R. F. Room-Temperature Bose–Einstein Condensation of Cavity Exciton–Polaritons in a Polymer. *Nature Mater* **2014**, *13* (3), 247–252. https://doi.org/10.1038/nmat3825.

(22) Chikkaraddy, R.; de Nijs, B.; Benz, F.; Barrow, S. J.; Scherman, O. A.; Rosta, E.; Demetriadou, A.; Fox, P.; Hess, O.; Baumberg, J. J. Single-Molecule Strong Coupling at Room Temperature in Plasmonic Nanocavities. *Nature* **2016**, *535* (7610), 127–130. https://doi.org/10.1038/nature17974.

(23) Santhosh, K.; Bitton, O.; Chuntonov, L.; Haran, G. Vacuum Rabi Splitting in a Plasmonic Cavity at the Single Quantum Emitter Limit. *Nat Commun* **2016**, *7* (1), ncomms11823. https://doi.org/10.1038/ncomms11823.

(24) Weber, T.; Kühner, L.; Sortino, L.; Ben Mhenni, A.; Wilson, N. P.; Kühne, J.; Finley, J. J.; Maier, S. A.; Tittl, A. Intrinsic Strong Light-Matter Coupling with Self-Hybridized Bound States in the Continuum in van Der Waals Metasurfaces. *Nat. Mater.* **2023**, *22* (8), 970–976. https://doi.org/10.1038/s41563-023-01580-7.

(25) Amo, A.; Lefrère, J.; Pigeon, S.; Adrados, C.; Ciuti, C.; Carusotto, I.; Houdré, R.; Giacobino, E.; Bramati, A. Superfluidity of Polaritons in Semiconductor Microcavities. *Nature Phys* **2009**, *5* (11), 805–810. https://doi.org/10.1038/nphys1364.
(26) Long, J. P.; Simpkins, B. S. Coherent Coupling between a Molecular Vibration and Fabry–Perot Optical Cavity to Give Hybridized States in the Strong Coupling Limit. *ACS Photonics* **2015**, *2* (1), 130–136. https://doi.org/10.1021/ph5003347.
(27) George, J.; Shalabney, A.; Hutchison, J. A.; Genet, C.; Ebbesen, T. W. Liquid-Phase Vibrational Strong Coupling. *J. Phys. Chem. Lett.* **2015**, *6* (6), 1027–1031. https://doi.org/10.1021/acs.jpclett.5b00204.
(28) Shalabney, A.; George, J.; Hutchison, J.; Pupillo, G.; Genet, C.; Ebbesen, T. W. Coherent Coupling of Molecular Resonators with a Microcavity Mode. *Nat Commun* **2015**, *6* (1), 5981. https://doi.org/10.1038/ncomms6981.
(29) Vergauwe, R. M. A.; George, J.; Chervy, T.; Hutchison, J. A.; Shalabney, A.; Torbeev, V. Y.; Ebbesen, T. W. Quantum Strong Coupling with Protein Vibrational Modes. *J. Phys. Chem. Lett.* **2016**, *7* (20), 4159–4164. https://doi.org/10.1021/acs.jpclett.6b01869.
(30) Memmi, H.; Benson, O.; Sadofev, S.; Kalusniak, S. Strong Coupling between Surface Plasmon Polaritons and Molecular Vibrations. *Phys. Rev. Lett.* **2017**, *118* (12), 126802. https://doi.org/10.1103/PhysRevLett.118.126802.
(31) Menghrajani, K. S.; Nash, G. R.; Barnes, W. L. Vibrational Strong Coupling with Surface Plasmons and the Presence of Surface Plasmon Stop Bands. *ACS Photonics* **2019**, *6* (8), 2110–2116. https://doi.org/10.1021/acsphotonics.9b00662.
(32) Yoo, D.; de León-Pérez, F.; Pelton, M.; Lee, I.-H.; Mohr, D. A.; Raschke, M. B.; Caldwell, J. D.; Martín-Moreno, L.; Oh, S.-H. Ultrastrong Plasmon–Phonon Coupling via Epsilon-near-Zero Nanocavities. *Nat. Photonics* **2021**, *15* (2), 125–130. https://doi.org/10.1038/s41566-020-00731-5.
(33) Hertzog, M.; Munkhbat, B.; Baranov, D.; Shegai, T.; Börjesson, K. Enhancing Vibrational Light–Matter Coupling Strength beyond the Molecular Concentration Limit Using Plasmonic Arrays. *Nano Lett.* **2021**, *21* (3), 1320–1326. https://doi.org/10.1021/acs.nanolett.0c04014.
(34) Dayal, G.; Morichika, I.; Ashihara, S. Vibrational Strong Coupling in Subwavelength Nanogap Patch Antenna at the Single Resonator Level. *J. Phys. Chem. Lett.* **2021**, *12* (12), 3171–3175. https://doi.org/10.1021/acs.jpclett.1c00081.
(35) Cohn, B.; Das, K.; Basu, A.; Chuntonov, L. Infrared Open Cavities for Strong Vibrational Coupling. *J. Phys. Chem. Lett.* **2021**, *12* (29), 7060–7066. https://doi.org/10.1021/acs.jpclett.1c01438.
(36) Muller, E. A.; Pollard, B.; Bechtel, H. A.; Adato, R.; Etezadi, D.; Altug, H.; Raschke, M. B. Nanoimaging and Control of Molecular Vibrations through Electromagnetically Induced Scattering Reaching the Strong Coupling Regime. *ACS Photonics* **2018**, *5* (9), 3594–3600. https://doi.org/10.1021/acsphotonics.8b00425.
(37) Sun, K.; Sun, M.; Cai, Y.; Levy, U.; Han, Z. Strong Coupling between Quasi-Bound States in the Continuum and Molecular Vibrations in the Mid-Infrared. *Nanophotonics* **2022**, *11* (18), 4221–4229. https://doi.org/10.1515/nanoph-2022-0311.
(38) Xie, P.; Cheng, Y. Manipulating Coherent Interaction of Molecular Vibrations with Quasibound States in the Continuum in All-Dielectric Metasurfaces. *Phys. Rev. B* **2023**, *108* (15), 155412. https://doi.org/10.1103/PhysRevB.108.155412.
(39) Xie, P.; Deng, Y.; Zeng, L.; Liang, Z.; Shen, S.; Ding, Q.; Zhang, H.; Zhou, Z.; Wang, W. Tunable Interactions of Quasibound States in the Continuum with Cavity Mode in a Metasurface-Microcavity Hybrid. *Phys. Rev. B* **2022**, *106* (16), 165408. https://doi.org/10.1103/PhysRevB.106.165408.
(40) Burnett, J. H.; Kaplan, S. G.; Stover, E.; Phenis, A. Refractive Index Measurements of Ge; LeVan, P. D., Sood, A. K., Wijewarnasuriya, P., D'Souza, A. I., Eds.; San Diego, California, United States, 2016; p 99740X. https://doi.org/10.1117/12.2237978.

(41) Li, H. H. Refractive Index of Alkaline Earth Halides and Its Wavelength and Temperature Derivatives. *Journal of Physical and Chemical Reference Data* **1980**, *9* (1), 161–290. https://doi.org/10.1063/1.555616.
(42) Baranov, D. G.; Wersäll, M.; Cuadra, J.; Antosiewicz, T. J.; Shegai, T. Novel Nanostructures and Materials for Strong Light–Matter Interactions. *ACS Photonics* **2018**, *5* (1), 24–42. https://doi.org/10.1021/acsphotonics.7b00674.
(43) Huang, L.; Krasnok, A.; Alú, A.; Yu, Y.; Neshev, D.; Miroshnichenko, A. E. Enhanced Light–Matter Interaction in Two-Dimensional Transition Metal Dichalcogenides. *Rep. Prog. Phys.* **2022**, *85* (4), 046401. https://doi.org/10.1088/1361-6633/ac45f9.
(44) Tavis, M.; Cummings, F. W. Exact Solution for an $N$-Molecule---Radiation-Field Hamiltonian. *Phys. Rev.* **1968**, *170* (2), 379–384. https://doi.org/10.1103/PhysRev.170.379.
(45) Saleh, B. E. A.; Teich, M. C. *Fundamentals of Photonics*; John Wiley & Sons, 2019.